# Strong variations of cosmic ray intensity during thunderstorms and associated pulsations of the geomagnetic field

K. Kh. Kanonidi[1], N. S. Khaerdinov[2], A. S. Lidvansky[2], and L. E. Sobisevich[3]

[1] Pushkov Institute of Terrestrial Magnetism, Ionosphere and Radio Wave Propagation, Russian Academy of Sciences, Troitsk, Moscow region, Russia

[2] Institute for Nuclear Research, Russian Academy of Sciences, Moscow, Russia

[3] Shmidt Institute of Physics of the Earth, Russian Academy of Sciences, Moscow, Russia

**Abstract**
Strong variations of the intensity of secondary cosmic rays during thunderstorms are found to be accompanied in some cases by very clear pulsations of the geomagnetic field. The experiment is carried out in the Baksan Valley, North Caucasus, the Carpet air shower array being used as a particle detector. Magnetic field measurements are made with high-precision magnetometers located deep underground in the tunnel of the Baksan Neutrino Observatory, several kilometers apart from the air shower array.

**Introduction**

When studying the intensity variations of secondary cosmic rays during thunderstorms [1] with the Carpet shower array of the Baksan Neutrino Observatory it was found that, in addition to regular variations correlating with the near-ground electric field, there existed considerable transient changes of the intensity [2, 3].

These sporadic variations are represented only by enhancements in the case of the soft component (electrons, positrons, and gamma rays). They can last a few minutes (as a rule, no longer than 10 min) reaching maximum amplitudes of excess over the background (daily mean) values of 20-30% [4]. For some enhancements of the soft component very sharp drops of increased intensity to the background level were observed simultaneously with lightning strokes. Therefore, for some time they were called pre-lightning enhancements [3, 5]. However, later it has become clear that enhancements in the soft component without any lightning effects are also observed. Such enhancements of the soft component intensity were also observed in some other experiments [9-13].

In the hard component (low-energy muons) also both regular variations and sporadic changes of intensity are observed during thunderstorms. In this case, there is a difference in the sign of effects: for muons both linear and quadratic terms in the regression dependence on the field intensity are negative [14], while for the soft component the quadratic coefficient is positive [2, 3]. As for sporadic variations of the muon intensity, they can be of two signs (positive and negative) and their amplitude never exceeds 1%, the characteristic time being equal to 8 min [16]. The event presented in this paper demonstrates an enhancement in the soft component and rather complex behavior of the hard component: multiple decreases of intensity and after that an enhancement almost coinciding with a similar effect in the soft component. In addition, approximately at the same time high-sensitivity magneto-variation station located deep underground several kilometers apart from the air shower array detected clear pulsations of the geomagnetic field. Some indications that thunderstorms can be a source of such pulsations were published previously (see, for example, [7]). The aim of this paper is to present direct experimental proof of excitation of the geomagnetic pulsations during thunderstorms and to draw attention to the fact that experimentally observed variations of secondary cosmic rays during thunderstorms can give a key to understanding the mechanism of excitation of these pulsations.

**Experiment**

The Carpet air shower array of the Baksan Neutrino Observatory includes 200 m$^2$ of scintillators under a concrete roof with a thickness of 29 g cm$^{-2}$. It is used as a single detector of particles with a threshold of 50 MeV corresponding to a mode energy release of relativistic penetrating particles. Taking into account ionization losses in the roof, this gives us the so-called hard component flux of secondary cosmic rays: mainly muons with energy threshold of about 100 MeV. The soft component of cosmic rays is measured by six huts with 9 m$^2$ of scintillators (54 m$^2$ in total) with only a very thin covering. For these uncovered scintillators two integral discriminators with threshold 10 and 30 MeV allowed one to isolate the soft component within these limits. Thus, we measured every second the hard component (E > 100 MeV, basically muons) and the soft component (10--30 MeV, electrons and gamma rays), the counting rates of the components being 40000 s$^{-1}$ and 4000 s$^{-1}$, respectively. In addition, the air shower array includes a muon detector (its total area and counting rate are 175 m$^2$ and 19000 s$^{-1}$, respectively) with an energy threshold of 1 GeV. The atmospheric pressure *P*, electric field strength *D*, and precipitation electric current *I* were also recorded every second. The electric field meter (of the electric mill type) is installed on the roof of the building where the Carpet of scintillators is located. The measuring electrode in this instrument is a two-sector rotating impeller connected with the ground through the load *R*. Above the electrode there is an immovable screen with corresponding sector cut-outs, which allow the electrode to be unscreened for measurements. The amplitude of the variable potential on the load (proportional to the acting electric field) is measured. In order to exclude the influence of charged rain droplets the instrument includes an umbrella located above the measuring electrode and rotating together with it. The speed of rotation and the size of the instrument were adjusted so that the measuring electrode would leave the open sector before the rain droplets in their free fall could cross the distance between the umbrella and the electrode and reach the latter.

In this paper we use also the data of magneto-variation station installed in Laboratory no.1 of North-Caucasus geophysical observatory. This laboratory is located in a separate underground cavity at a distance of 4100 m from the entrance into the tunnel of the Baksan Neutrino Observatory. The three-component digital magneto-variation station is installed on a concrete base in the chamber's center, being oriented along the magnetic meridian. Three quartz magnetic sensors of the Bobrov's type continuously measure three components (h, d, and z) of the magnetic induction vector. The frequency range of measurements is from 0 to 1 Hz, the random error at measurements lasting no less than 3 seconds is 0.1 nT. Relative time instability of the instrument is 1-2 nT per year, and the dynamic range of measurements is $\pm$ 2000 nT.

**The event on October 15, 2007**

Thunderstorm on October 15, 2007 in Baksan Valley lasted for 24 hours. More exactly, it was raining during this period, while the thunderstorm proper (strong electric activity) consisted of three episodes. During the last of them a strong disturbance of the soft component of secondary cosmic rays took place, as one can see in Fig. 1 (the second panel from the top). This disturbance represents an enhancement with a maximum of 3.5% approximately from 18: 35 to 18:45 of local time, the statistical error for the soft component being equal to 0.3%. In the hard component approximately at the same time an enhancement is also observed. Its maximum exceeds 0.8% at a statistical error of 0.1% (the second panel from the bottom in Fig. 2). This peak of muon intensity is preceded by two decreases of intensity of approximately the same amplitude and duration (more exactly, duration of one of them is twice longer, but probably it can be doubled, and there are in fact three rather than two decreases).

Generally, the event does not look unusual. As far as the amplitudes of observed effects are concerned, it is even rather modest. Here, the maximum enhancement of the soft component hardly exceeds 5% (at the best time resolution), while in the events published earlier the maximum enhancements are bigger than 20% (September 7, 2000 [1, 2]) and 30% (October 11, 2003 [4]).

However, the remarkable feature of the event under consideration is the existence of pulsations of the geomagnetic field. Figure 1 presents all three components of the magnetic field measured at a large distance from the air shower array (about 4 km) and under a huge overburden of rock. One can clearly see pulsations lasting for about 40 min. However, their amplitude is small in comparison with the daily wave. Therefore, in the bottom panel of Fig. 2 one component of the geomagnetic field (h-component) is presented with subtracted daily trend. In this case it becomes clear that, in addition to pulsations with a period of about 100 s, there is also a slower and longer variation with a period of about 1 h. The form of slow pulsations in the period of approximately from 16:45 to 18:20 coincides with the profile of muon variations, though the latter are delayed with respect to the magnetic field variation by 9 min.

**Discussion and conclusions**

Pulsations of the geomagnetic field are a widely known and comparatively well studied phenomenon [14]. According to commonly accepted classification [15] they occur to be of two basic types (Pc и Pi), and the type of pulsations is indicated by one of the above pairs of letters with a subsequent digit that identifies the frequency of pulsations. Origination of the pulsations is ascribed to generation of Alfvenic or magnetosonic waves in the magnetosphere, and sharp changes in the solar wind density and velocity are considered as their source, as well as interplanetary shock waves. Roughly speaking, classical geomagnetic pulsations are a result of knocking on the magnetosphere from the side of outer space.

The data presented in this paper allow one to suggest that approximately the same effect can be caused by the processes occurring in the stratosphere during thunderstorms. It was already said above that some indications to magnetic pulsations during thunderstorms had been obtained in paper [7][1].

Considerable variations of muon intensity (~ 1%) originate when the top of a thunderstorm cloud is elevated substantially higher than the effective level of muon production, since only in this case there is a sufficiently strong resulting field for mouns propagating from the point of generation to the observation level. In the region between the cloud top and ionosphere a large potential difference (~ 200 MV) is formed [16]. This field is higher than the critical field for generation of runaway electron avalanches, which should lead to mass ionization of the stratosphere in the acceleration region. Due to polarization of the ionized stratosphere, powerful atmospheric currents originate. Apparently, they are a source of excitation of the slow pulsations to which one can ascribe the type Pi3 (T > 150 s). These slow pulsations demonstrate the time behavior that is very similar to the time profile of the muon intensity, though there is a time delay between them equal to about 9 min.

Unlike the slow pulsations, fast pulsations have no analogs in the profiles of cosmic ray variations: the latter do not show disturbances with a characteristic time of 100 s. However, these fast pulsations are clearly recorded by magnetic sensors as pulses on the background of a slow positive disturbance (Fig. 2). And at the same time we observe prolonged continuous disturbances in both soft and hard components of secondary cosmic rays. One can suggest that the flux of particles generated in the region of strong filed is bidirectional (or an overlapping of two simultaneous processes takes place): first, it produces some effect on the observation level;

---

[1] However, in some other papers (see, for example, [17]) the search for magnetic pulsations during thunderstorms gave a negative result.

second, rather strong flux of particles should reach the magnetosphere to excite classical geomagnetic pulsation of Pc4 or Pi2 type (T = 40-150 s). In the last case, nonlinear and resonance processes can take place, as is typical for geomagnetic pulsations. However, the available experimental data are so far insufficient to construct a particular model for this scenario. Nevertheless, we cannot but indicate to the possibility of this mechanism looking at the coincidence of experimental data of different nature in Figs 1 and 2. The probability of the cause-and-effect connection between them should be very high in this event.

**Acknowledgments**

The work is supported by the Russian Foundation for Basis Research (grant nos. 08-02-00613 and 09-02-00274).

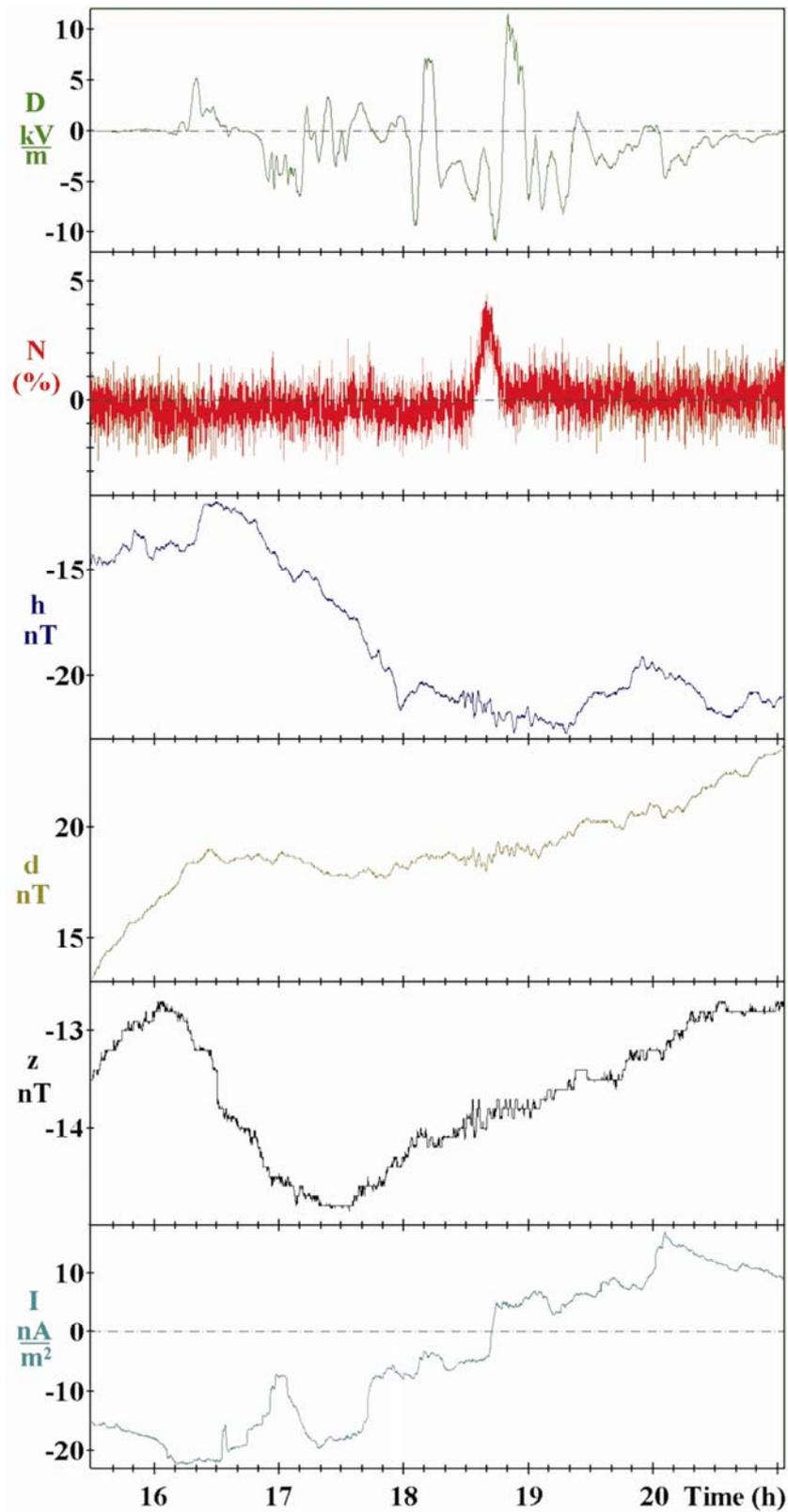

**Figure 1.** Thunderstorm event observed in Baksan Valley on October 15, 2007. Interval of averaging is 4 s. From top to bottom the panels represent the strength of the near-earth electric field, intensity of the soft component 10-30 MeV (percent deviation from daily mean value), three components of the geomagnetic field, and precipitation electric current. Abscissa axis presents local time (UTC + 3 h).

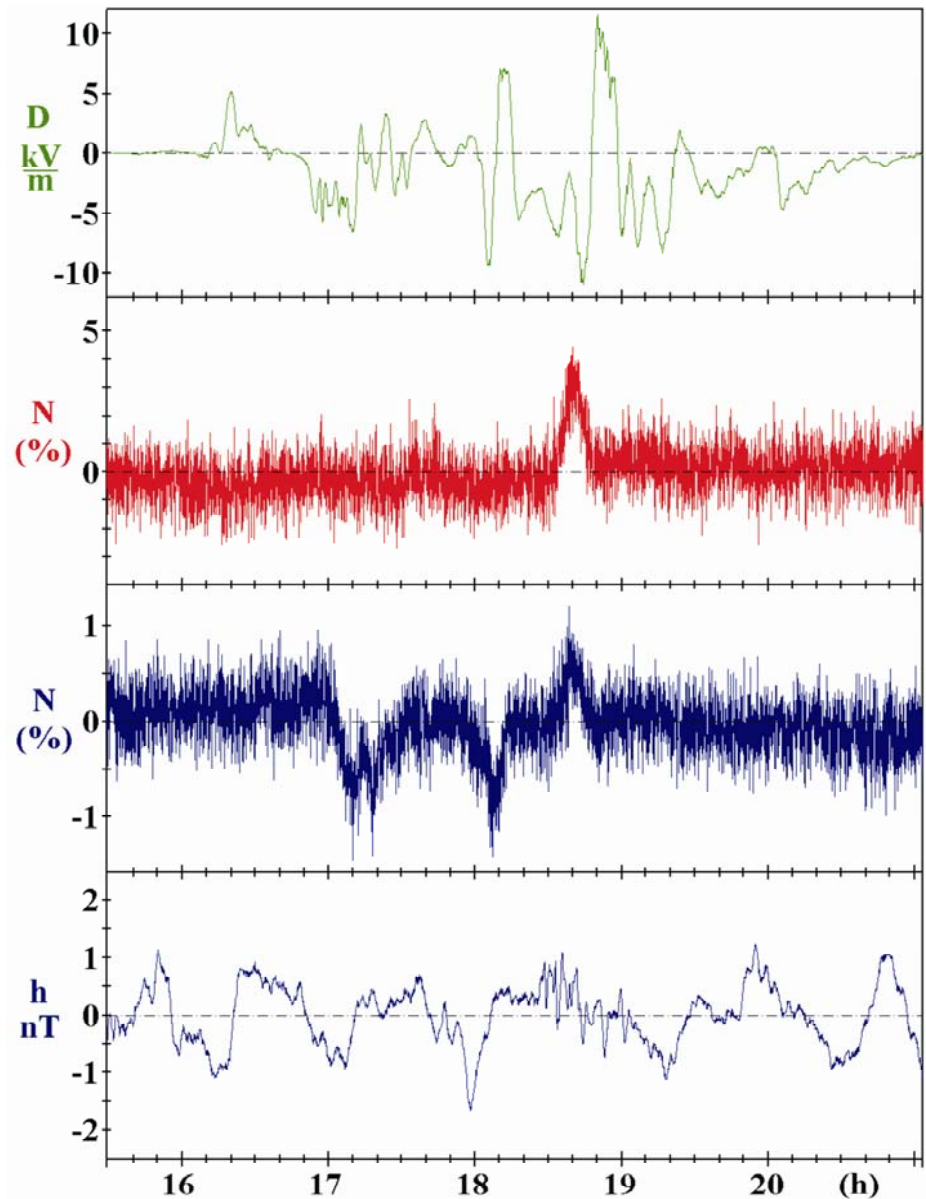

**Figure 2.** Two upper panels repeat the data of Fig. 1. The third one presents the intensity of muons with energy threshold of about 100 MeV, and the bottom panel shows the h-component of the geomagnetic field with subtracted trend of the daily wave. One can clearly see periodicity with a period of about 1 h. Pulsations with characteristic period of about 100 s are also observed near the time of simultaneous enhancements of the soft and hard components. Suppressions in the muon intensity repeat the profile of magnetic field variation with a delay of about 9 min. Abscissa axis presents local time (UTC + 3 h).